\documentclass{article}
\usepackage{graphicx} 
\usepackage{amsmath}
\usepackage{booktabs} 
\usepackage{makecell}

\title{Mapping Academic Sports Analytics:\\
A Content Analysis of JQAS and JSA\thanks{This work was supported by the Israel Science Foundation [Grant 2718/25].}}
\author{Adam Nathaniel\thanks{Department of Statistics and Operations Research, School of Mathematical Sciences, Tel Aviv University. e-mail: adamnathaniel6@gmail.com.}}
\date{2026}

\begin{document}

\maketitle

\begin{abstract}
Sports analytics has expanded rapidly in professional sports and mainstream culture, yet defining its academic core remains a challenge. In this paper, we provide an empirical map of the academic sports analytics landscape based on a systematic content analysis of approximately 700 papers published in two leading dedicated journals, the \textit{Journal of Quantitative Analysis in Sports} (JQAS) and the \textit{Journal of Sports Analytics} (JSA), between the years 2005 and 2024. We classify each paper by sport discipline, mathematical method, and core research area, identify recurring themes, and examine macro trends. We find game tactics, player evaluation, team ranking, result forecasting, and tournament design to be the most studied research areas in sports analytics. We highlight selected papers in each area, illustrating the wide range of mathematical methods used and sports disciplines analyzed. Our work offers a comprehensive overview of the questions, methods, and disciplines that characterize the field, serving as a foundational reference for practitioners and researchers in sports analytics.
\end{abstract}

\section{Introduction}

Sports analytics has increasingly been gaining popularity and momentum in mainstream culture in recent years, ever since the release of the movie {\it Moneyball} in 2011. It has also become an integral part of the sports industry, as many sports clubs now hire data scientists for analytics tasks and maintain close ties with sports-tech companies. 

A natural follow-up question concerns the role of sports analytics in academia. What exactly constitutes the academic field of sports analytics? As this is a relatively new area of study, the answer is not entirely straightforward. Szymanski (2020) even compared the current state of the field with alchemy rather than science due to its limited reliance on theory. This paper aims to help answer that question, drawing inspiration from Wright (2009) and Swartz (2020).

Wright (2009) reviewed roughly 50 years of research in Operations Research (OR) as applied to sports. The review did not attempt to provide a complete history, but rather to give the reader a taste of what is now a large and diverse field. It identified the main areas of research in OR in sports and discussed numerous papers in each area, showing the wide range of mathematical methods used and sports disciplines analyzed.

Swartz (2020) discussed issues relevant to publishing in sports analytics. One of the primary difficulties in getting a handle on sports analytics research is that it is widely distributed. Sports analytics papers often exist in journals as one-off papers, and there are relatively few pure sports analytics journals. The {\it Journal of Quantitative Analysis in Sports} (JQAS), founded in 2005, was considered by Swartz (2020) to be the premier outlet for sports analytics research. The {\it Journal of Sports Analytics} (JSA), founded in 2015, was also highlighted as a rapidly gaining recognition.

Both journals address sports analytics, but each has a different focus. JQAS emphasizes technical and methodological innovations, while JSA prioritizes practical sporting contributions. As a result, these journals complement each other, providing a comprehensive view of both the theoretical and applied aspects of the field. For this study, we reviewed 20 volumes and more than 500 papers from JQAS, together with 10 volumes and more than 150 papers from JSA. 

However, the primary purpose of this paper is not to be a survey, but rather to map the academic sports analytics landscape through an empirical content analysis of its core literature. From Wright we learn what to look for and from Swartz we learn where to look for it. By cataloging and categorizing nearly 700 papers from JQAS and JSA, we identify recurring research areas in sports analytics, and highlight selected papers in each area. In addition, to provide a broader perspective, we discuss several prominent sports analytics surveys in the following section.

We classify each paper in both journals by its mathematical method, sport discipline, and core research area. Among the various topics in sports analytics, game tactics appear to be the most studied. Other central themes include player evaluation, team ranking, result forecasting, and tournament design. 

The range of mathematical frameworks applied in sports analytics is extensive. However, some methods are more commonly used than others. Naturally, machine learning and deep learning models serve as leading tools in forecasting results. Frequently used frameworks for analyzing game tactics include game theory and causal inference. These domains represent complementary approaches: game theory provides a normative framework for strategic behavior, and causal inference addresses empirical questions of cause and effect.

Popular techniques include Markov chains, Monte Carlo simulations, Bayesian statistics, and various types of regression models. Certain domains are also closely tied to specific sports disciplines, such as extreme value theory, which is often used to analyze performance in track and field events. In addition, we review methods developed specifically for sports analytics, such as the Elo rating system.

The journals cover more than 50 sports disciplines. In the early days, baseball was the discipline most analyzed, but today soccer is leading the way. Basketball and American football are also frequently studied, both at the professional and college levels. Similarly, tennis, golf, cricket, and ice hockey have received considerable attention.

The literature also features papers on traditional sports, including track and field disciplines and Olympic events such as the decathlon. Alongside them there are disciplines like chess and eSports, whose status as sports is occasionally debated. Women's sports are also being discussed, particularly in the areas of volleyball and tennis.

The categorization of papers into distinct research areas in subsequent sections inevitably involves a degree of subjectivity. To introduce a more objective dimension to this meta-analysis, we examine keywords that recur in multiple papers. The most popular sports are at the forefront: {\it baseball, basketball, football,} and {\it soccer}. The most prevalent mathematical methodologies and challenges, {\it simulation}, {\it machine learning}, {\it ranking}, and {\it prediction}, along with the overarching terms {\it sports} and {\it sports analytics} themselves, follow closely. A more detailed list of recurring keywords is provided in the appendix.

The remainder of this paper is organized as follows. Section 2 reviews related work and historical surveys. Section 3 outlines our corpus and provides a macro-level content analysis across publication periods, sporting disciplines, and quantitative methodologies. Sections 4 through 8 examine the five core research domains: game tactics (Section 4), player evaluation (Section 5), team ranking (Section 6), result forecasting (Section 7), and tournament design (Section 8), detailing their quantitative footprints alongside selected papers. Section 9 addresses additional research areas that emerge from our study, and Section 10 concludes the paper.

\section{Related Work}

To the best of our knowledge, our work of reviewing around 700 papers from JQAS and JSA and identifying core research areas is unique. However, there are various other sports analytics surveys, each with a different focus. In this section, we review some of the main surveys and meta-analysis papers related to sports analytics. 

Gerchak (1994) reviewed the development of operations research and management science (OR/MS) in sports dating back to the 1950s. The paper noted that sports provide a natural setting for these fields, being structured and goal-oriented. It placed particular emphasis on post-1976 literature, directing readers to Machol and Ladany (1976) and Ladany and Machol (1977) for a comprehensive bibliography of earlier work. The main research areas reviewed by Gerchak (1994) were strategies in team sports and decision making in individual sports. Tactical questions were examined in disciplines such as baseball and American football, along with broader structural issues like player drafts in North American leagues and tournament scheduling problems. In his concluding remarks, Gerchak (1994) contrasted the optimism of the late 1960s and mid-1970s regarding OR/MS in sports with the more skeptical perspective of the mid-1990s, observing that sports franchises seem to over-collect data and then under-analyze them.

OR in sports was also the topic of Wright (2009). Like Gerchak (1994), Wright (2009) traced the origins of the field to the 1950s, identifying Mottley (1954) as the first paper to consider implementing scientific methods to improve decision-making in sports, while also noting the work done by Machol and Ladany in the late 1970s. It then argued that most published papers in the field are not highly theoretical or mathematical, a perspective later echoed by Szymanski (2020), and that the literature in the field lacks coherence because many studies are "one-off" in nature, an observation similar to the one in Swartz (2020). The paper also noted that American researchers tend to study American sports, such as baseball, American football, and basketball, while non-American researchers put more focus on other sports such as soccer and cricket. The main research areas of OR in sports, according to Wright (2009), are tactics and strategy, scheduling, and forecasting. Tactics represent one of the first domains explored, both regarding decisions that need to be taken in real time, such as substitution strategies, and decisions that need to be taken between matches or seasons, such as replacing the team manager. Scheduling research addresses problems including the scheduling of fixtures and officials, while forecasting focuses primarily on predicting match outcomes, an area closely linked to the sports gambling industry.

Interestingly, while scheduling was mentioned as an integral part of OR in sports by Gerchak (1994) and Wright (2009), Swartz (2020) placed it on the periphery of sports analytics, arguing that it is less concerned with how a sport is actually played. Another field closely related to sports analytics is sports science. The {\it Journal of Sports Sciences} published a 25th Anniversary Edition featuring a survey of the journal's papers on sports performance (Nevill et al., 2008). The sports performance section of the journal includes mathematical and statistical research. Among its core research domains, according to Nevill et al. (2008), are talent identification and the relative age effect, home advantage and its influence on officials, and the analysis of fatigue and injuries in training.

Morgulev et al. (2018) highlighted the modern explosion of sports data, a phenomenon already observed by Gerchak (1994) and later characterized by Szymanski (2020) as a fundamental aspect of sports analytics today. Morgulev et al. (2018) reviewed key technological developments in notational analysis (the systematic and objective recording and tracking of performance data) and identified the mid-1990s as the beginning of the big-data era in sports. It demonstrated the use of big data and data-driven analysis in sports by providing representative examples of research, both performance-oriented and business-oriented. In addition, it presented how the analysis of sports data enables us to learn about human behavior, a perspective also emphasized by Bar-Eli et al. (2020). 

The explosion of sports data was also central in Davis et al. (2024), which highlighted the applications of machine learning in team sports in the areas of player recruitment, match preparation, and management of a player's health and fitness. Davis et al. (2024) discussed the use of different data types (physical data, match sheet data, event data, and tracking data) and addressed the unique methodological challenges inherent in modeling sports data.

Referring to the big-data era in sports, Szymanski (2020) compared the current state of sports analytics to alchemy rather than science, as there is a lot of effort and data invested in the field, generating valuable discoveries, but with limited reliance on theory. This theoretical gap must be addressed to allow sports analytics to make a significant contribution to our understanding of sports. Furthermore, the paper advocated for greater openness and transparency within the discipline. With this in mind, Szymanski (2020) highlighted the Sports Performance Analytics MOOC developed by the University of Michigan, which teaches sports analytics skills using Python. Following an introductory module using sports data to teach general data analysis concepts, the course covers the mathematics behind Moneyball, develops predictive and machine learning models to forecast match outcomes, and presents wearable technologies used in sports science. 

For instance, the underlying theory of Moneyball is explained based on Hakes and Sauer (2006). In short, drawing a walk was historically an undervalued skill in baseball. The on-base percentage (OBP, which values hits and walks) explains winning better than the slugging percentage (SLG, which values hits but not walks). However, prior to the publication of Moneyball (Lewis, 2003), SLG explained salaries better than OBP. This market inefficiency allowed Billy Beane, the general manager of the Oakland A's, to recruit undervalued players using data-driven analysis. Another example presented in the course is an ordinal logistic regression model used to predict the winner of a soccer match based on the ratio of Transfermarkt values between the home and away teams. These values are freely available online, estimated according to the wisdom of the crowd methodology, and the results of this simple prediction procedure are not significantly inferior to those of state-of-the-art sophisticated models.

More recently, Baumer et al. (2023) presented four key ideas in sports analytics. First, the estimation of the expected value of a game-state at any given point, representing the expected number of points that the team will score over some determined amount of time. Second, win probability, the estimate of the probability that a team will win the game, given its current state. Third, the estimation of team strength, based on win-loss records, pairwise comparisons, and other methods. Fourth, analysis of data from sports betting markets, which are considered one of the best sources for estimating the true probability of events. A fifth concept, the individual player rating and evaluation for team sports, is also briefly mentioned. Like our work, Baumer et al. (2023) aimed to provide an overview for anyone interested in sports analytics.

Several academic books on sports analytics have also been published over the years, each with a different emphasis in terms of research questions, target disciplines, and quantitative methodologies. Although there is no single definitive text in the field, Swartz (2020) described the {\it Handbook of Statistical Analyses in Sports} edited by Albert et al. (2017) as the most current snapshot of research in sports analytics. From an educational standpoint, Elmore and Urbaczewski (2025) authored the {\it Introduction to Sports Analytics using R}, with the aim of providing the first textbook in the field. Both books organized their coverage around specific disciplines, dedicating individual chapters to major sports, including baseball, American football, basketball, ice hockey, and soccer.

\section{Corpus Overview}

Our corpus includes 687 papers. Of them, 517 are from the first 20 volumes of JQAS (2005-2024) and 170 are from the first 10 volumes of JSA (2015-2024). We classify each paper in both journals by its sport discipline, mathematical method, and core research area. Importantly, we initially reviewed the literature without imposing an a-priori taxonomy and subsequently identified recurring research questions that formed the basis of the classification. In this section, we present the macro insights of our content analysis. In order to evaluate macro trends, we further group the papers in blocks of five years per journal. The distribution of the papers over these blocks is presented in Table~\ref{tab:paper_counts}.

\begin{table}[htbp]
\centering
\small
\caption{Number of Published Papers Across Five-Year Eras.}
\label{tab:paper_counts}
\begin{tabular}{lcccccc}
\toprule
& \textbf{2005--2009} & \textbf{2010--2014} & \multicolumn{2}{c}{\textbf{2015--2019}} & \multicolumn{2}{c}{\textbf{2020--2024}} \\
\cmidrule(lr){4-5} \cmidrule(lr){6-7}
& \textbf{JQAS} & \textbf{JQAS} & \textbf{JQAS} & \textbf{JSA} & \textbf{JQAS} & \textbf{JSA} \\
\midrule
Papers & 114 & 209 & 85 & 81 & 109 & 89 \\
\bottomrule
\end{tabular}
\end{table}


\subsection{Sport Disciplines}

The disciplines most commonly studied include soccer (with 106 papers analyzing it as their main sport discipline), basketball (97), baseball (84) and American football (78), with around $60\%$ of all papers focusing on one of them. The second tier of disciplines, with around $20\%$ together, includes tennis (30), ice hockey (29), golf (29), cricket (28) and volleyball (15). More than 50 other disciplines are studied. However, fewer than ten papers focus on each of them.

\begin{figure}[htbp]
    \centering
    \includegraphics[width=1\textwidth]{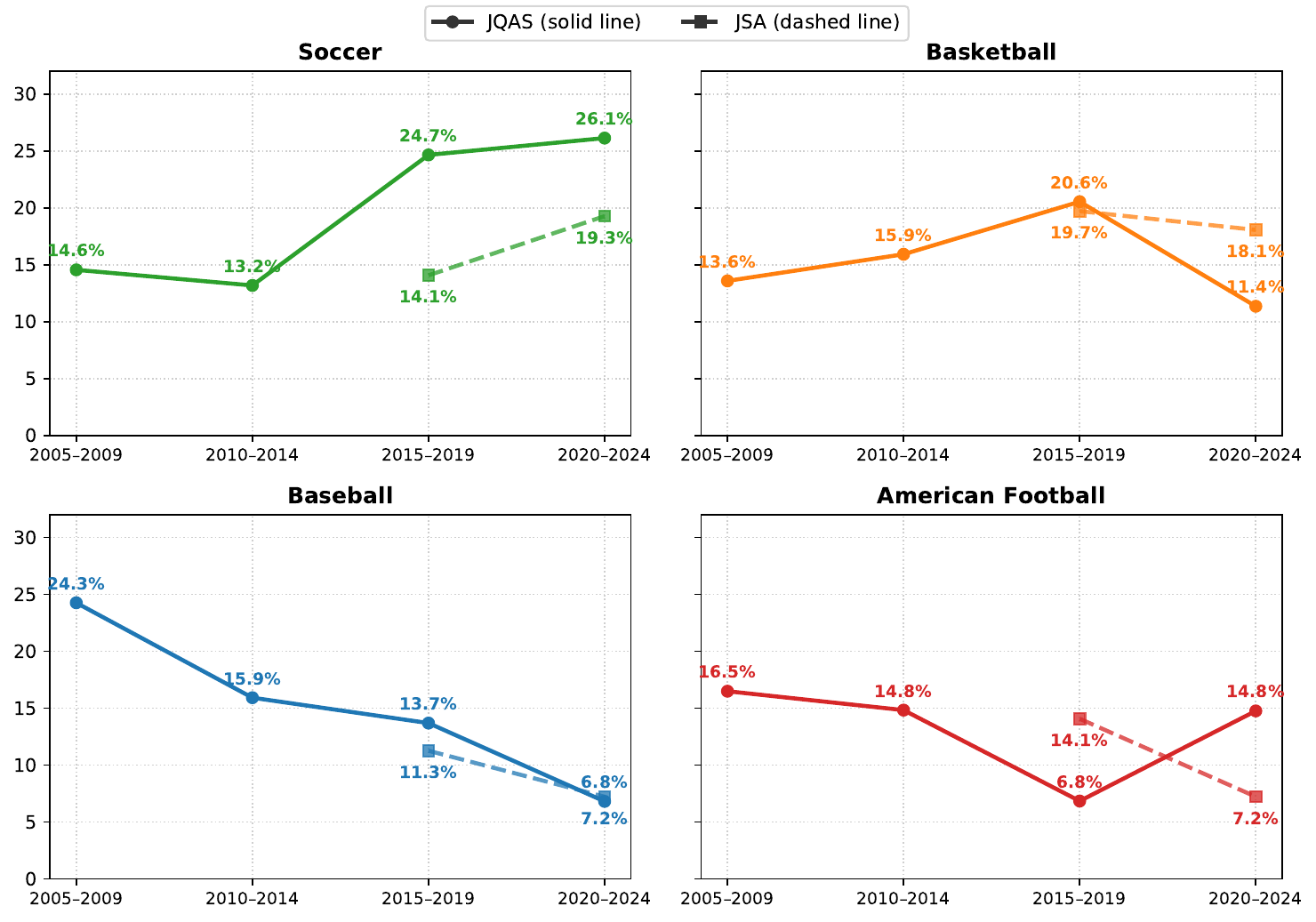}
    \caption{Proportion of Papers per Sport Across Five-Year Eras.}
    \label{fig:sports_proportions_plot}
\end{figure}

The share of each of the leading sports (soccer, basketball, baseball, and American football) over time is summarized in Table~\ref{tab:sports_proportions} and visualized in Figure~\ref{fig:sports_proportions_plot}. It emphasizes that, while baseball was the discipline most analyzed in the past, its popularity has decreased over time, whereas the popularity of soccer has steadily increased in both journals. In fact, if we combine the two blocks of 2020-2024, the number of baseball papers is similar to that of cricket and tennis.

\begin{table}[htbp]
\centering
\small
\setlength{\tabcolsep}{2.99pt} 
\caption{Proportion of Papers per Sport Across Five-Year Eras.}
\label{tab:sports_proportions}
\begin{tabular}{lcccccc}
\toprule
& \textbf{2005--2009} & \textbf{2010--2014} & \multicolumn{2}{c}{\textbf{2015--2019}} & \multicolumn{2}{c}{\textbf{2020--2024}} \\
\cmidrule(lr){4-5} \cmidrule(lr){6-7}
& \textbf{JQAS} & \textbf{JQAS} & \textbf{JQAS} & \textbf{JSA} & \textbf{JQAS} & \textbf{JSA} \\
\midrule
Soccer & 14.56\% & 13.19\% & 24.66\% & 14.08\% & 26.14\% & 19.28\% \\
Basketball & 13.59\% & 15.93\% & 20.55\% & 19.72\% & 11.36\% & 18.07\% \\
Baseball & 24.27\% & 15.93\% & 13.70\% & 11.27\% & 6.82\% & 7.23\% \\
American Football & 16.50\% & 14.84\% & 6.85\% & 14.08\% & 14.77\% & 7.23\% \\
\midrule
\textbf{Total (Top 4)} & \textbf{68.92\%} & \textbf{59.89\%} & \textbf{65.76\%} & \textbf{59.15\%} & \textbf{59.09\%} & \textbf{51.81\%} \\
\bottomrule
\end{tabular}
\end{table}

Although the distribution of disciplines is informative on its own, a deeper analysis reveals further patterns. For example, we examine the distributions of women's and college sports. Approximately $6\%$ of the papers focus on women's sports, with more than $30\%$ of them analyzing volleyball and tennis. Moreover, half of the papers that study volleyball analyze women's volleyball. The distribution of college sports and NCAA papers is also unique. Approximately $11\%$ of the papers in our corpus study college sports. Of them, more than $70\%$ analyze basketball and American football.

\subsection{Mathematical Methods}

Unlike classifying sports disciplines, categorizing mathematical methods is far less clear-cut. A paper can use many mathematical methods at the same time, but it can also use no specific method at all. Moreover, we may distinguish between mathematical frameworks and quantitative tools. While the first group includes a wide range of methodologies to describe the environment of the problem, the latter focuses on methods used to model specific processes. Due to the high potential for ambiguity in this discussion, we mostly highlight recurring fields rather than providing a rigid classification.
 
According to our analysis, many of the papers in the journals mainly use standard statistics. This may support the claim of Wright (2009) that most papers in the field are not highly theoretical or mathematical, and the claim of Szymanski (2020) that the field's reliance on theory nowadays is limited. It is further supported by Wakelam et al. (2022), who conducted a survey on the attributes of soccer players and individual player performance. The survey included 132 papers from a wide range of publishers, including the {\it Journal of Performance Analysis in Sport}, the {\it Journal of Sports Sciences}, the {\it MIT Sloan Sports Analytics Conference}, and JQAS. Wakelam et al. (2022) found that more than half of the 132 papers were solely applying descriptive statistical techniques.

In general, there is a large variety of mathematical methods in the journals. However, the most popular technique is regression models (mainly logistic and linear regressions), with approximately $10\%$ to $15\%$ of all papers using them. Markov chains and Monte Carlo simulations have also been frequently used in the past, while today there is an increasing implementation of Bayesian statistics, especially in JQAS, as they appear in approximately $20\%$ of the journal's papers in the years 2015-2024. The volume of spatial statistics papers has also been growing consistently over time, perhaps due to the increased availability of tracking data (Rahimian and Toka, 2022). The popularity of the Elo rating system has also recently grown in both journals, while other frequently used ranking methods (such as Bradley-Terry, Massey, Colley, and PageRank) did not experience a similar phenomenon.

\begin{figure}[htbp]
    \centering
    \includegraphics[width=0.75\textwidth]{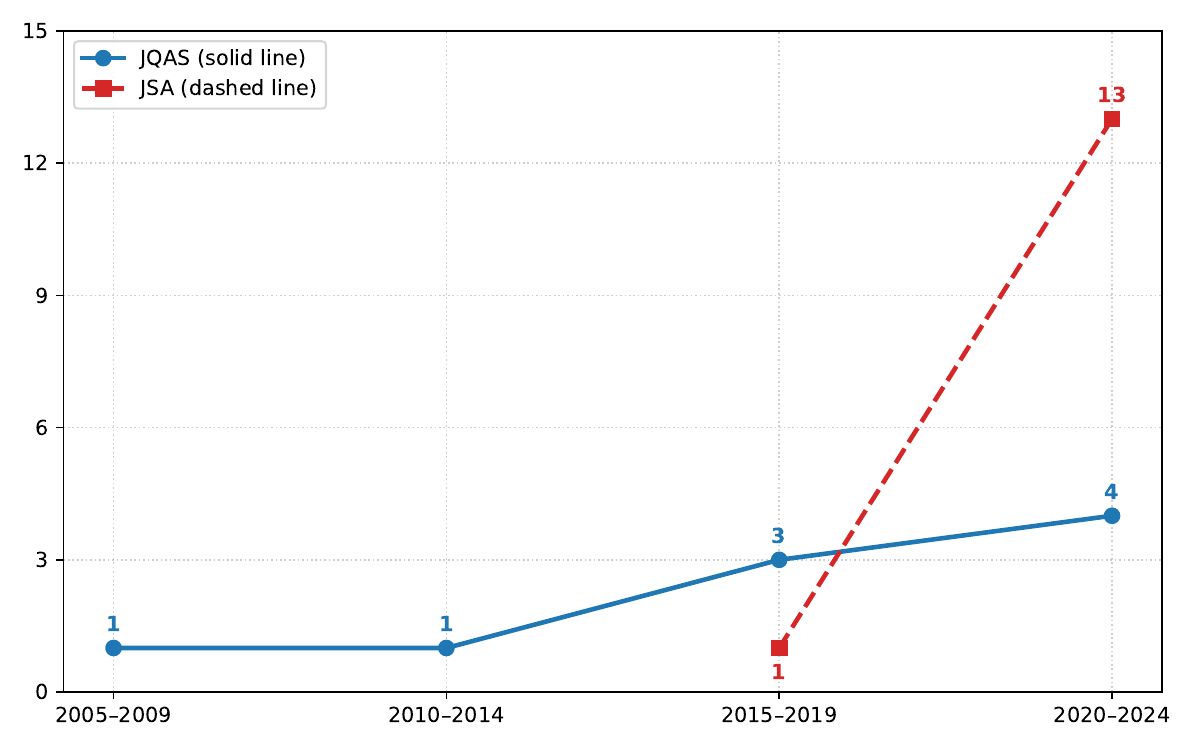}
    \caption{Number of Papers with Machine Learning Keyword Across Five-Year Eras.}
    \label{fig:machine_learning_plot}
\end{figure}

In terms of the popularity of classic mathematical domains, the leading field is machine learning, followed by deep learning, game theory, and causal inference. Other recurring mathematical frameworks include time series analysis, survival analysis, integer programming, and extreme value theory. As we will discuss in detail later, some fields are mainly used in specific research areas, such as machine learning for result forecasting, game theory and causal inference for game tactics, and the Elo rating system for team ranking. Although leading, the majority of machine learning papers (according to both our analysis and an objective count of keywords) have been published over the years 2020-2024 in JSA (with a relatively balanced distribution of papers over this time range), as can be seen in Figure~\ref{fig:machine_learning_plot}.

\subsection{Research Areas}

Among the various topics in sports analytics, we find five leading research areas. Game tactics appears to be the most studied topic, with 79 papers in our corpus. Other central themes include player evaluation (71 papers), team ranking (62), result forecasting (50), and tournament design (44). Each of these five topics is examined in a dedicated section, with analysis of macro trends and presentation of selected papers. The main research question of approximately $40\%$ to $50\%$ of all papers in the journals belongs to one of these five research areas. Another $20\%$ to $25\%$ of the papers belong to one of a few smaller recurring research areas, presented in Section 9, while the rest of the papers are "one-off" papers.

\section{Game Tactics}

Approximately $10\%$ to $15\%$ of papers published in the journals focus on game tactics and optimizing in-game decisions. The disciplines with the highest relative share of game tactics papers out of all their papers include American football, soccer, baseball, and basketball. Almost $18\%$ of all American football papers deal with tactical questions, and another $6\%$ focus specifically on play calling. More than $20\%$ of all tactics papers analyze soccer. Around $25\%$ of all tactics papers implement game theory, causal inference, machine learning, or deep learning. About $80\%$ of all game theory papers analyze tactics, and so do approximately $25\%$ of all Markov chain papers. The keywords {\it causal inference}, {\it strategy}, {\it player tracking}, {\it tactics}, and {\it game theory} appear almost exclusively in tactics papers. Finally, in the years 2020-2024, approximately $50\%$ of all tactics papers in JQAS used tracking data. In what follows, we present a few selected game tactics papers.

Annis (2006) introduced a simple decision tree model to address the long-debated basketball strategy of whether a team leading by three points should intentionally foul the opponent in the final seconds of a game. The paper compared conventional defensive strategies with intentional fouling and concluded that, given reasonable estimates of relevant probabilities, the latter strategy is more likely to lead to the defensive team’s victory.

Hirotsu and Wright (2006) applied a game-theoretic approach to analyze tactical formation changes in soccer. They assumed that the number of goals scored is a Poisson random variable with intensity that depends on real match data. The study also assumed that the offensive and defensive strengths of each formation are independent of the opponent's formation, and modeled the interaction between the formations of two teams as a non-cooperative zero-sum game. The paper derived recursive differential formulas that computed the winning probabilities depending on the changes in formations, and identified optimal substitution strategies by considering both the score and the remaining time in the match. A non-zero-sum model was presented in Hirotsu et al. (2009).

Kopriva and Vecer (2009) examined the impact of red cards in soccer using data from the minute-by-minute betting market. By modeling the number of goals scored by each team as a Poisson distribution, they derived team scoring rates from betting odds, taking into account the current game score and remaining time. This approach, combined with the granularity of minute-by-minute betting data, enabled them to quantify changes in scoring rates during matches as a result of red cards. By understanding how red cards influence scoring rates, they identified scenarios where committing a red card offense may be strategically advantageous to prevent a goal-scoring opportunity.

In scoring-based games, teams do not necessarily aim to maximize their total score but rather to maximize the probability of outscoring their opponents. Skinner (2011) analyzed this concept mathematically, modeling the game as a sequence of independent plays, so that the final score follows a binomial distribution and can be approximated by a Gaussian distribution using the central limit theorem. Consequently, in certain situations, it may be optimal for teams to select plays with a lower expected score but a higher variance in scoring outcomes. The study presented theoretical examples, such as the decision between attempting two-point or three-point shots in basketball, which depends on factors such as the number of remaining possessions and the team's current deficit.

Crossing the ball has long been a common tactic in soccer. However, its effectiveness can be questioned. Wu et al. (2021) address this issue using a spatio-temporal analysis enabled by the recent availability of tracking data, which refers to the $(x, y)$ coordinates of the ball and the 22 players recorded at frequent time intervals. Using observational data, they aimed to estimate the causal effect of crossing (treatment) on shooting opportunities (outcome) by applying propensity score matching with covariates associated with game situations. Specifically, they define crossing opportunities as instances in which the offensive team retains possession in a wide area within the attacking third of the pitch. To model the propensity score, the probability that a player decides to cross, they employed logistic regression based on spatio-temporal factors, such as the ratio of offensive to defensive players in the box. They matched each crossing opportunity that resulted in a cross with a similar opportunity, in terms of propensity score, that did not result in a cross, and then compared the outcomes of these matched observations. Their findings indicated that the average effect of treatment on the treated is positive, suggesting that, in general, players who choose to cross the ball are making an effective decision.

Biro and Walker (2022) modeled an American football possession as a Markov decision process and applied a reinforcement learning algorithm, specifically value iteration, to optimize play calling. The states were defined by the current down, yards to the first down, and yards to the goal line, while the action space consisted of running and passing. The reward mechanism was based on maximization of expected points through possession, and transition probabilities were empirically estimated from league-wide data. Individual player or team data were not incorporated, as this would require an exponentially larger amount of data. A special case corresponding to the final five minutes of the game was also considered, in which additional information, namely time remaining and score differential, was incorporated into the state space, and the objective shifted from maximizing expected points to maximizing win probability.

In other research, Barnett et al. (2010) applied game theory and contest theory to analyze a best-of-three match between two players in which players decide whether to increase their effort. Hirotsu (2011) found a batting order in baseball that outperforms the order that maximizes the expected number of runs, by also considering the standard deviation of runs and the probability of winning the match. Myers (2012) and Silva and Swartz (2016) debated substitution strategies in soccer. Nadimpalli and Hasenbein (2013) used a Markov decision process to analyze when a player should challenge a line call in tennis to optimize the probability of winning, based on the context of the match and the confidence of the player that the call was wrong. Bauer et al. (2023) identified and analyzed team formation in soccer, using tracking data and deep learning.

\section{Player Evaluation}

Approximately $10\%$ of papers published in the journals focus on player evaluation in team sports. The disciplines with the highest relative share of player evaluation papers out of all their papers include baseball, ice hockey, cricket, and American football. Almost $18\%$ of all baseball papers deal with player evaluation questions, and more than $20\%$ of all player evaluation papers analyze baseball. Keywords associated with player evaluation papers include {\it adjusted plus-minus}, {\it player evaluation}, and {\it random effects}. Finally, regression models and Bayesian statistics are each employed in approximately $20\%$ of all player evaluation papers. In what follows, we present a few selected player evaluation papers.

Gelade (2014) introduced a post-shot expected-goals metric in soccer, which estimates the probability that a shot-on-target results in a goal, given the observed trajectory of the ball. A logistic regression model was trained to predict the binary outcome of each shot (goal vs. no goal), using shots-on-target data from two seasons in the English Premier League. The explanatory variables included representations of the shot start-point (by distance from the goal line and distance from the pitch midline) and end-point (by distance from the goal midline and height of the ball). The model fitted the data well and allowed evaluation of goalkeeper performance. Compared to the saves-to-shots ratio, which assumes that all shots are equally difficult to stop, the new metric was more stable per goalkeeper across seasons and avoided the systematic bias in favor of goalkeepers of stronger teams, who tend to face easier shots.

Bransen et al. (2019) developed a framework to evaluate soccer players by quantifying the contribution of their passes to the expected goals of their teams. The expected added value of a pass was defined as the difference between its expected end value and its expected start value. The expected start value was computed as the average value of all possession sequences that include the starting location of the pass, where a possession was assigned a value of zero if it did not end with a shot, and otherwise the expected-goals value of the resulting shot. The expected end value was estimated by considering passes that are similar to the pass under evaluation, using a domain-specific distance function that captured both the characteristics of the pass and contextual factors. It was calculated as a weighted average of the values of the possession sequences following these similar passes. Due to the absence of a clear ground truth for the value of passes, the authors evaluated the proposed metric indirectly and showed that it predicted future team performance, as measured by match outcomes, from past performances more accurately than commonly used pass evaluation metrics.

In other research, Franks et al. (2016) developed meta-metrics to quantify and compare metrics that are used to assess player ability. Wulff and De Silva (2022) used Pareto optimality to evaluate baseball players based on multiple criteria, as a replacement to weighted aggregation. Matano et al. (2023) developed an adjusted plus-minus model to have a one-number statistics to evaluate soccer players, with FIFA video game ratings as a Bayesian prior. Baron et al. (2024) introduced an improved post-shot expected-goals metric in soccer, in which non-zero values are assigned to off-target shots. Schutz and Tofallis (2024) suggested a new scoring method for decathlon athletes, based on the concept of Pareto efficiency.

\section{Team Ranking}

Approximately $10\%$ of papers published in JQAS in the years 2005-2014 focused on team ranking, and their ratio decreased and became closer to $5\%$ in both journals in the years 2015-2024. The disciplines with the highest relative share of team ranking papers out of all their papers include American football and baseball. Almost $18\%$ of all American football papers deal with team ranking questions, and more than $30\%$ of all team ranking papers analyze American football. Naturally, most papers employing the Elo ranking system focus on team ranking questions, as do papers utilizing paired comparisons or models such as Bradley-Terry, Thurstone-Mosteller, Massey, Colley, and PageRank. Keywords associated with team ranking papers include {\it ranking}, {\it sports rankings}, {\it ratings}, {\it rating systems}, {\it paired comparisons}, {\it elo}, {\it elo algorithm}, {\it elo ratings}, and {\it NCAA basketball}. Finally, approximately $25\%$ of the papers analyzing the "March Madness" tournament ask questions related to team ranking. In what follows, we present a few selected team ranking papers.

Paired comparison models are used to estimate the relative strengths of competing entities based on head-to-head matches. Linear models for paired comparisons assume that team \( i \) has a "playing strength" represented by a single real number \( v_i \), and that the probability of team \( i \) defeating team \( j \) can be expressed as the cumulative distribution function (CDF) of the difference in their strengths: \( \pi(v_i, v_j) = H(v_i - v_j) \). Two of the most common models are the Bradley-Terry model, where \( H(x) = \frac{\exp(x)}{1 + \exp(x)} \), corresponding to the CDF of the logistic distribution, and the Thurstone-Mosteller model, where \( H(x) = \Phi(x) \), the CDF of the standard normal distribution. Chan (2011) established "linearity conditions" under which a function \( \pi(i, j) \) can be expressed as \( H(v(i) - v(j)) \) for some CDF \( H \) and an increasing function \( v \), which is unique up to a linear transformation. For example, the function \( \pi(i, j) = \frac{i}{i + j} \) satisfies these conditions and can be expressed in the Bradley-Terry framework with \( v(a) = \ln(a) \). The paper then examined three simple theoretical games that almost always satisfy the linearity conditions, in order to evaluate the predictive accuracy of the Bradley-Terry and Thurstone-Mosteller models. To do so, it introduced an algorithm for defining the function \( v \). The results indicated that both models achieve high predictive accuracy, with the Thurstone-Mosteller model, generally the less popular of the two, performing slightly better.

In other research, Constantinou and Fenton (2013) suggested a dynamic rating system for soccer teams based on match results, compared it with Elo, and used it as a betting strategy. Szczecinski and Djebbi (2020) showed the connection between Elo and Bradley-Terry models, and derived an extended Elo-Davidson algorithm for sports with a draw result. Devlin et al. (2021) developed an iterative rating method and compared it with Elo and Bradley-Terry. Szczecinski (2022) derived an extended G-Elo algorithm for sports with a goal difference or a margin of victory aspect in result.

\section{Result Forecasting}

Approximately $10\%$ to $15\%$ of papers published in the journals in the years 2015-2024 focus on outcome prediction and result forecasting, while it was a less common research area in the years 2005-2014. The disciplines with the highest relative share of result forecasting papers out of all their papers include tennis, cricket, basketball, and ice hockey. More than $25\%$ of all result forecasting papers analyze basketball, although biased by a special issue dedicated to prediction methods in basketball (Glickman and Sonas, 2015). About $30\%$ of all result forecasting papers implement machine learning or deep learning, while other papers use regression models and Elo-based predictions. Keywords associated with result forecasting papers include {\it prediction}, {\it machine learning}, {\it sports forecasting}, {\it maximum likelihood}, and {\it betting}. Finally, approximately $25\%$ of the papers analyzing the "March Madness" tournament ask questions related to result forecasting. In what follows, we present a few selected result forecasting papers.

David et al. (2011) introduced new concepts for predicting the outcomes of American football games using artificial neural networks (ANNs). Among them, one approach involved structuring input variables by averaging opposing but corresponding team statistics. For example, they calculated the mean of the points scored by a team per game and the points conceded by its opponent per game to capture a balanced statistical differential. Another approach was the "committee of committees" method, which builds on the "committee of machines" strategy. In this framework, multiple ANNs are trained on different partitions of the random data set, and their predictions are aggregated. By further combining the outputs of multiple such committees, the "committee of committees" aims to refine prediction accuracy. Their approach achieved accuracy comparable to the Las Vegas bookmakers lines and outperformed many other computer-based NFL prediction models.

Lock and Nettleton (2014) proposed a random forest model for estimating the probability of a win at any point during an NFL game, using twelve seasons of play-by-play data. Explanatory variables included pre-game betting odds, current score, time remaining, and situational factors such as down and yards-to-go. In addition, the authors introduced an adjusted score variable defined as $\text{AdjustedScore}(\gamma) = \frac{\text{Score}}{(\text{Seconds}+1)^\gamma}$, with the parameter $\gamma = 0.5$ selected on the basis of cross-validation. Predictive accuracy, measured by mean squared error from all plays, increases as the game progresses, with betting odds emerging as the most informative predictor in the first quarter, and score becoming dominant in subsequent quarters. Changes in predicted win probability can be used to identify the most influential plays and to assess tactical decisions, although the authors noted the limitations of using observational rather than experimental data. To evaluate the model, plays in the test set were grouped into bins based on their estimated win probabilities, with the actual win rate in each bin compared to the average predicted probability, yielding extremely high calibration. The methodology is generalizable and can be adapted to other sports contexts.

In other research, Braunstein (2010) extended the Pythagorean formula for estimating winning percentages in baseball by incorporating performance consistency. Kovalchik (2016) compared the performance of 11 prediction models in tennis, including Elo-based models. Wheatcroft (2021) used a two-step idea to predict match results in soccer by first predicting match statistics.

\section{Tournament Design}

Approximately $5\%$ to $10\%$ of papers published in the journals focus on tournament design and tournament structure. The disciplines with the highest relative share of tournament design papers out of all their papers include golf and soccer. More than $20\%$ of all tournament design papers analyze soccer. Keywords associated with tournament design papers include {\it tournament design}, {\it draw procedure}, {\it sports scheduling}, {\it OR in sports}, {\it fairness}, {\it balance}, and {\it simulation}. Finally, about $18\%$ of all tournament design papers use simulations. In what follows, we present a few selected tournament design papers.

Eugster et al. (2010) challenged the widely held belief that playing the second leg at home provides an advantage in soccer knockout ties, using data from the UEFA Champions League knockout phase. While a naive analysis supports this belief, a more rigorous examination reveals a different conclusion. Analyzing matches in which the second leg's venue was randomly assigned showed a balanced 50/50 chance of progression. Furthermore, the authors used a logistic regression model across the entire dataset, accounting for factors such as group stage performance (group winner or runner-up) and team strength, as estimated by the UEFA team coefficient. Their analysis did not find statistical evidence to support the conjecture that having the second leg at home offers a significant advantage.

Guyon (2018) analyzed tournament structures with 24 teams, consisting of a group stage with six groups of four teams followed by a 16-team knockout stage, motivated by the format of UEFA EURO 2016, in which the four best third-placed teams advance to the knockout phase. The paper highlighted several disadvantages of this structure, including an inherent arbitrary group advantage (as some group winners do not face another group winner until the semifinals) and a lack of win incentives (since in some groups it is not always optimal to finish the group stage in the highest possible position). The author then proposed and analyzed an alternative format, which ranks the 16 qualified teams in a manner analogous to the existing third-placed team ranking: group winners are ranked 1 to 6, group runners-up 7 to 12, and the best third-placed teams 13 to 16. Given these rankings, a standard knockout bracket can be employed, although such brackets have their own drawbacks in the absence of reseeding.

In other research, Baumann et al. (2009) and Zimmerman and Lim (2021) discussed the middle seed anomaly, which suggests that in 16-team seeded knockout formats (such as March Madness sweet 16), it is better to be seeded 10/11 than 8/9. Puterman and Wang (2011) studied the effect of a relegation-promotion system on league competitiveness. Guyon (2015, 2021) examined the flaws in the FIFA World Cup formats over the years, in terms of fairness, competitive balance, and risk of collusion, and suggested improved formats. Dagaev and Rudyak (2019) simulated and analyzed the effect of the reform of the seeding system in the UEFA Champions League.

\section{Other Research Areas}

As we mentioned, approximately $20\%$ to $25\%$ of all papers in the journals belong to one of a few small recurring research areas. The biggest areas, out of these small areas, deal with home advantage, officiating, competitive balance and fairness, records, age, draft, strategy (off-game decisions, in contrast to tactics), handicap, scouting, competitions randomness, hot hand and momentum, hall of fame, play calling, and injuries. Some of these areas can be seen as sub-areas of the big five, such as play calling for game tactics and competitive balance for tournament design. March Madness is also a recurrent theme, although not exactly a research area, but rather appears in other research areas, such as team ranking and result forecasting. In addition, approximately $30\%$ to $35\%$ of all papers are "one-off" papers.

For instance, Arkes (2010) investigated the hot hand phenomenon in the context of basketball free throws. Specifically, the study tested whether making the first free throw increases the probability of making the second. However, a straightforward analysis would be biased because better shooters are naturally more likely to make both throws. To address this, the author used a regression model that combined all players, using an indicator variable for the outcome of the previous free throw and controlling for player-specific fixed effects in a multivariate framework. Using data from an entire NBA season, the results indicated that successfully making the prior free throw increased the probability of making the next one by 2 to 3 percentage points. The study also included a brief discussion on the potential for a "cold hand" effect.

\section{Conclusion and Discussion}

In this paper, we provided an empirical map of the academic sports analytics literature based on a systematic review and classification of approximately 700 papers published in two leading dedicated journals, JQAS and JSA. We classified each paper by sport discipline, mathematical method, and core research area. We reviewed the literature without imposing an a-priori taxonomy but rather identified recurring research questions that formed the basis of the classification. 

We found game tactics, player evaluation, team ranking, result forecasting, and tournament design to be the most studied research areas in the field. Since the categorization of papers into distinct research areas inevitably involves a degree of subjectivity, we also examined keywords that recur in multiple papers, to introduce an objective dimension to our meta-analysis. Our work may serve as an overview for anyone interested in sports analytics, including new researchers in the field.

Throughout the paper, we focused on the academic field of sports analytics. However, its origins are firmly grounded in industry practices. While this can raise challenges, such as a limited reliance on theory, highlighted by Szymanski (2020), it can also be beneficial. In this last discussion section, we briefly mention three examples of what we view as a valuable combination between academic and industrial sports analytics.

The most cited paper in both journals, by a significant margin, is Kubatko et al. (2007), with over 500 citations. It brought together key insights from the quantitative analysis of basketball that were widely accepted in the industry, providing a peer-reviewed foundation for future academic research. It discussed central concepts in basketball analytics, such as equal possessions for opponents in a match. It also presented established metrics such as efficiency, alongside practical guidelines such as avoiding per-minute statistics for players with fewer than 500 or 1,000 minutes in a season. 

The high impact of Kubatko et al. (2007) in the academic literature emphasizes the academic value of industrial insights in general. Similarly, a place that reflects the industrial value of academic insights is the {\it MIT Sloan Sports Analytics Conference}. Being the most famous sports analytics venue, it provides opportunities for academic researchers to engage with cutting-edge developments and industry perspectives, and present their research to senior executives in sports clubs.

Finally, we present an example of the mutual benefit from this academic-industry connection, which may also affect mainstream culture. As mentioned before, there is no single definitive text in academic sports analytics, while the most famous industrial sports analytics book is {\it Moneyball}, published in 2003, in which the journalist Michael Lewis described the sabermetric approach of the general manager Billy Beane. We claim that its potential soccer successor, {\it How to Win the Premier League} (Graham, 2024), enjoyed the academic-industry duality of sports analytics.

Ian Graham, a Physics PhD, served as Liverpool FC director of research in the years 2012-2023 and developed one of the first in-house analytics departments in the English Premier League. In {\it How to Win the Premier League}, he revealed the data revolution that helped Liverpool win their first league title in 30 years. Since it is written by its main protagonist, the book describes Graham's data-driven and evidence-based approach in depth. In addition, Graham highlighted three academic papers that were essential to his work: Pollard and Reep (1997), which introduced early notions of expected goals and expected threat to evaluate players, alongside Dixon and Coles (1997) and Dixon and Robinson (1998), which modeled team strength to forecast match outcomes. We find {\it How to Win the Premier League} to be a compelling example of the balance between mathematical theory, practical implementation, and a bestselling story.

\newpage
\appendix
\section{Recurring Keywords}
The following keywords recurred five or more times in the papers we reviewed (after combining different words with similar meanings, such as "soccer" and "association football", or "Prediction" and "Predictions").

\begin{table}[htbp]
\centering
\label{tab:keyword_frequencies_symmetrical}
\small
\begin{tabular*}{\textwidth}{@{\extracolsep{\fill}}lcclcc@{}}
\hline
\textbf{Keyword} & \textbf{Count} & & \textbf{Keyword} & \textbf{Count} \\
\hline
Baseball & 49 & & Probability & 8 \\
Basketball & 47 & & Ratings & 8 \\
Football & 45 & & Strategy & 8 \\
Soccer & 42 & & Tournament & 8 \\
Simulation & 30 & & American Football & 7 \\
Sports & 24 & & Bayesian Hierarchical Model & 7 \\
Machine Learning & 23 & & Bayesian Modeling & 7 \\
Ranking & 23 & & Cluster Analysis & 7 \\
Prediction & 21 & & Integer Programming & 7 \\
Sports Analytics & 20 & & Major League Baseball & 7 \\
Performance & 19 & & Random Forests & 7 \\
Tennis & 19 & & Rating Systems & 7 \\
Golf & 18 & & Sports Performance & 7 \\
Markov Chain & 18 & & Sports Statistics & 7 \\
NFL & 18 & & Team Performance & 7 \\
NHL & 15 & & Tournament Design & 7 \\
Logistic Regression & 14 & & ANOVA & 6 \\
NBA & 14 & & Bayesian Models & 6 \\
Competitive Balance & 13 & & Bias & 6 \\
Bayesian & 12 & & Elo & 6 \\
Cricket & 12 & & Estimation & 6 \\
Hockey & 12 & & Fairness & 6 \\
Home Advantage & 12 & & March Madness & 6 \\
Optimization & 12 & & Olympics & 6 \\
College Football & 11 & & Pythagorean Method & 6 \\
MCMC & 11 & & Adjusted Plus-minus & 5 \\
NCAA & 11 & & Causal Inference & 5 \\
Statistics & 11 & & Clustering & 5 \\
Game Theory & 10 & & College Basketball & 5 \\
Sports Forecasting & 10 & & Elo Ratings & 5 \\
Volleyball & 10 & & Lasso & 5 \\
Win Probability & 10 & & Monte Carlo & 5 \\
Classification & 9 & & Order Statistics & 5 \\
English Premier League & 9 & & PGA Tour & 5 \\
Paired Comparisons & 9 & & Pitching & 5 \\
Regression & 9 & & Survival Analysis & 5 \\
Analytics & 8 & & Variable Selection & 5 \\
\hline
\end{tabular*}
\end{table}

\end{document}